\documentstyle[preprint,aps]{revtex}
\begin{document}

\draft
\title{Natural Inflation From Fermion Loops}
\author{William H. Kinney\thanks{kinney@colorado.edu} and K.T.
Mahanthappa\thanks{ktm@verb.colorado.edu}}
\address{University of Colorado, Boulder}
\author{COLO-HEP-354, hep-ph/9503331 (revised)}
\date{August 16, 1995}
\maketitle

\begin{abstract}
``Natural'' inflationary theories are a class of models in which inflation is
driven by a pseudo-Nambu-Goldstone boson. In this paper we consider two
models, one old and one new, in which the potential for inflation is
generated by loop effects from a fermion sector which explicitly breaks a
global $U(1)$ symmetry. In both models, we retrieve the ``standard'' natural
inflation potential, $V\left(\theta\right) = \Lambda^4\left[1 +
\cos\left(\theta / \mu\right)\right]$, as a limiting case of the exact
one-loop potential, but we carry out a general analysis of the models
including the limiting case. Constraints from the COBE DMR observation and
from theoretical consistency are used to limit the parameters of the models,
and successful inflation occurs without the necessity of fine-tuning the
parameters.
\end{abstract}

\pacs{}

\section {Introduction}
\label{intro}

Inflationary cosmologies consider an epoch in the very early universe in
which the energy density of the universe was dominated by vacuum energy,
resulting in a period of exponential expansion of the
universe\cite{guth81,linde82,albrect82}. This vacuum energy is typically
generated in particle physics models by a scalar field potential associated
with a broken symmetry.
One attractive class of inflationary models is ``Natural'' inflation, which
uses the vacuum energy associated with a pseudo-Nambu-Goldstone boson (PNGB)
to drive inflation\cite{freese90,adams93}. In this paper, we consider two
models in which the PNGB is the result of a fermion sector which explicitly
breaks a global $U(1)$ symmetry: one earlier model involving chiral
fermions\cite{adams93} and a new model involving non-chiral fermions. Section
\ref{slowrollreview} gives a review of ``slow-roll'' inflationary models with
particular application to natural inflation. Section \ref{models} discusses
the two models in detail, and Section \ref{conclusions} presents conclusions.
We apply three conditions to the models discussed: (i) the model must satisfy
standard observational constraints on inflationary theories, particularly
those from the COBE DMR observation, (ii) the model must be weakly coupled,
so that perturbation theory is valid, and (iii) the PNGB must acquire a
nonzero mass at the minimum of its potential. These conditions allow us to
place constraints on the parameters of the models, and we find that succesful
inflation occurs without the necessity of fine-tuning. In both models, we
find that the inflationary potential used in Refs.\cite{freese90,adams93}
arises as a limiting case of the exact one-loop potential.

\section {Slow-Roll and Natural Inflation}
\label{slowrollreview}

Inflationary cosmologies explain the observed flatness and homogeneity of the
universe by postulating the existence of an epoch during which the energy
density of the universe was dominated by vacuum energy, resulting in a period
of exponential increase in the scale factor of the universe
\begin{equation}a(t) \propto e^{H t}\end{equation}
The {\it Hubble parameter} H is given by
\begin{equation} H^2 \equiv \left({\dot a \over a}\right)^2 = {8 \pi \over 3
m_{pl}^2} \rho_{vac} \simeq const.\end{equation}
where $m_{pl} \sim 10^{19} \mbox{GeV}$ is the Planck mass. Nonzero vacuum
energy is introduced into particle physics models by including a scalar field
$\phi$, the {\it inflaton}, with a potential $V\left(\phi\right)$. During
inflation, the inflaton is displaced from the minimum of its potential,
resulting in a nonzero vacuum energy, and evolves to the minimum with
equation of motion
\begin{equation}\ddot\phi + 3 H \dot\phi + V'\left(\phi\right) =
0\end{equation}
So-called ``new'' inflationary models consider potentials which contain at
least one region flat enough that the evolution of the field is friction
dominated
\begin{equation}3 H \dot\phi + V'\left(\phi\right) =
0\label{slowroll}\end{equation}
This is known as the {\it slow-roll} approximation. This approximation can be
shown to be valid if the {\it slow-roll parameters} $\epsilon$ and
$\left|\eta\right|$\cite{copeland93} are both much less than 1, where:
\begin{equation}\epsilon\left(\phi\right) \equiv {m_{pl}^2 \over 16 \pi}
\left[{V'\left(\phi\right) \over V\left(\phi\right)}\right]^2\end{equation}
\begin{equation}\eta\left(\phi\right) \equiv {m_{pl}^2 \over 8 \pi}
\left[{V''\left(\phi\right) \over V\left(\phi\right)} - {1 \over 2}
\left[{V'\left(\phi\right) \over
V\left(\phi\right)}\right]^2\right]\end{equation}
For an inflationary theory to correspond to the observed universe, it must at
minimum satisfy two conditions: (i) The scale factor of the universe must
increase by a factor of at least $e^{60}$ during the inflationary epoch, in
order to  explain the observed thermal equilibrium of the cosmic microwave
background radiation (CMB). (ii) Quantum fluctuations in the inflaton field
give rise to primordial density fluctuations, with calculable amplitude
$\delta$ and spectral index $n_s$\cite{hawking82,starob82,guth82,bardeen83},
which can be compared to the COBE DMR observation $\delta \sim 10^{-5}$, $n_s
= 1.1 \pm 0.5$\cite{smoot92,wright92}.
These conditions can be used to place constraints on the form of the
potential. Consider a scalar field initially at some value $\phi$. The field
evolves according to equation (\ref{slowroll}) to the minimum of the
potential, where it oscillates and decays into other particles ({\it
reheating}). Slow-roll ends and reheating commences at a field value $\phi_f$
where the slow-roll parameter $\epsilon\left(\phi_f\right)$ is
unity\cite{copeland93}:
\begin{equation}\epsilon\left(\phi_f\right) \equiv {m_{pl}^2 \over 16 \pi}
\left[{V'\left(\phi_f\right) \over V\left(\phi_f\right)}\right]^2 =
1\end{equation}
where $\epsilon\left(\phi\right) < 1$ during the slow-roll period. The number
of e-folds of inflation which occur when the field evolves from $\phi$ to
$\phi_f$ is
\begin{equation}N\left(\phi\right) = {8 \pi \over m_{pl}^2}
\int_{\phi_f}^{\phi}{{V\left(\phi'\right) \over
V'\left(\phi'\right)}\,d\phi'}\label{numefolds}\end{equation}
We determine the upper limit on the initial field value $\phi \leq \phi_i$
for which sufficient inflation occurs by requiring $N\left(\phi_i\right) =
60$. Quantum fluctuations in the inflaton field when $\phi \sim \phi_i$
produce density fluctuations on scales of current astrophysical interest. We
can constrain the potential by requiring that the magnitude of the density
fluctuations $\delta\left(\phi_i\right)$ produced during the inflationary
period match the observation by COBE\cite{stewart93}:
\begin{equation}\delta\left(\phi_i\right) = \left({2 \over \pi}\right)^{1 /
2} {\left[V\left(\phi_i\right)\right]^{3/2} \over m_{pl}^3
V'\left(\phi_i\right)} \left[1 - \epsilon\left(\phi_i\right) + \left(2 - \ln2
- \gamma\right) \left(2 \epsilon\left(\phi_i\right) -
\eta\left(\phi_i\right)\right)\right] \sim
10^{-5}\label{deltaCOBE}\end{equation}
where $\gamma \simeq 0.577$ is Euler's constant. In addition, it is possible
to calculate the spectral index $n_s$ of the density fluctuations. The
fluctuation power per logarithmic interval $P\left(k\right)$ is defined in
terms of the density fluctuation amplitude $\delta_k$ on a scale $k$ as
$P\left(k\right) \equiv |\delta_k|^2$. The {\it spectral index} $n_s$ is
defined by assuming a simple power-law dependence of $P\left(k\right)$ on k,
$P\left(k\right) \propto k^{n_s}$.
The spectral index $n_s$ of density fluctuations is given in terms of the
slow-roll parameters $\epsilon$ and $\eta$\cite{stewart93}:
\begin{equation}n_s \simeq 1 - 4 \epsilon\left(\phi_i\right) + 2
\eta\left(\phi_i\right)\label{scalarindex}\end{equation}
For $\epsilon,\,\left|\eta\right| \ll 1$ during inflation, inflationary
theories predict a nearly scale-invariant power spectrum, $n_s \sim 1$.
The value of the spectral index derived from the first year of COBE data is
$n_s = 1.1 \pm 0.5$\cite{smoot92,wright92}. The COBE two year results are
also available\cite{bennett94}. However, different statistical methods used
in analyzing the data lead to different bounds on the spectral
index\cite{bennett95}. For the purposes of this paper, we will take $n_s \geq
0.6$.

``Natural'' inflationary theories use a pseudo-Nambu-Goldstone boson to drive
the inflationary expansion. The basic scenario consists of the following:
A spontaneous symmetry breaking phase transition occurs at a scale $\mu$, and
temperature $T \sim \mu$. In this paper we will be considering symmetry
breaking involving a single complex scalar field $\phi$, with a potential
\begin{equation}V\left(\phi\right) = \lambda \left[\phi^* \phi -
\mu^2\right]^2\end{equation}
which is symmetric under a global $U(1)$ transformation $\phi \rightarrow
e^{i \alpha} \phi$.
At the minimum of the potential $V\left(\phi\right)$, we can parameterize the
scalar field as $\phi = \sigma \exp\left[{i \theta} / \mu\right]$.
The radial field $\sigma$ has a mass $M^2_\sigma \sim \lambda \mu^2$.
The field $\theta$ is a Nambu-Goldstone boson, and is massless at tree level.
If the $U(1)$ symmetry of the potential $V\left(\theta\right)$ is preserved
by the rest of the Lagrangian, $\theta$ will remain massless with loop
corrections. But if the $U(1)$ is broken by other terms in the Lagrangian,
$\theta$ acquires a potential $V_1\left(\theta\right)$ from loop corrections,
leading to a nonzero mass, and is called a {\it pseudo-Nambu-Goldstone boson}
(PNGB).
Assuming the mass of $\theta$ is much less than that of the radial mode,
$M^2_\theta \ll M^2_\sigma$, the field $\theta$ will be effectively massless
near the original symmetry breaking scale $T \sim \mu$. As the temperature of
the universe decreases, $T \ll M_\sigma$, excitations of the heavy $\sigma$
field will be damped, so that we can take $\sigma = \mu = const.$ The only
remaining degree of freedom will be $\theta$, and we can parameterize $\phi$
as:
\begin{equation}\phi = \mu e^{i {\theta \over
\mu}}\label{adiabatic}\end{equation}
At temperatures $T \gg M_\theta$, the effective potential
$V_1\left(\theta\right)$ is negligible. When the universe cools to $T \sim
M_\theta$, $V_1\left(\theta\right)$ becomes important\cite{freese90}.
The field $\theta$ will roll down the potential to its minimum, resulting in
inflationary expansion during the period in which the energy density of the
universe is dominated by vacuum energy. Natural inflationary models typically
assume a potential for the PNGB of the form
\begin{equation}V_1\left(\theta\right) = \Lambda^4 \left[1 +
\cos\left({\theta \over \mu}\right)\right]\label{stdnat}\end{equation}
This is a limiting case of the exact one-loop potential of the models
considered in this paper. $\mu$ is the original symmetry breaking scale, and
$\Lambda$ is an independent energy scale characterizing the temperature at
which the potential $V_1\left(\theta\right)$ becomes significant.

We can use (\ref{numefolds}) and (\ref{deltaCOBE}) to constrain the
parameters $\mu$ and $\Lambda$ as follows\cite{freese90}.
Inflation ends at a field value $\theta_f$ where the slow-roll parameter
$\epsilon\left(\theta_f\right) = 1$:
\begin{equation}\left({\theta_f \over \mu}\right) = \cos^{-1} \left[{1 - 16
\pi \left(\mu / m_{pl}\right)^2 \over 1 + 16 \pi \left(\mu /
m_{pl}\right)^2}\right]\end{equation}
The number of e-folds for a given initial field value $\theta_i$
(\ref{numefolds}) is independent of the scale $\Lambda$, but depends on the
scale $\mu$. Requiring $N\left(\theta_i\right) \geq 60$ results in an upper
bound $\theta \leq \theta_i$ in terms of $\mu$:
\begin{equation}\left({\theta_i \over \mu}\right) = \cos^{-1} \left[1 - 2 {16
\pi \left({\mu \over m_{pl}}\right)^2 \over 1 + 16 \pi \left({\mu \over
m_{pl}}\right)^2} \exp\left[-{60 \over 8 \pi} \left({m_{pl} \over
\mu}\right)^2\right]\right]\label{thetaupper}\end{equation}
so that $\theta_i$ decreases rapidly with decreasing scale $\mu$.

\begin{figure}
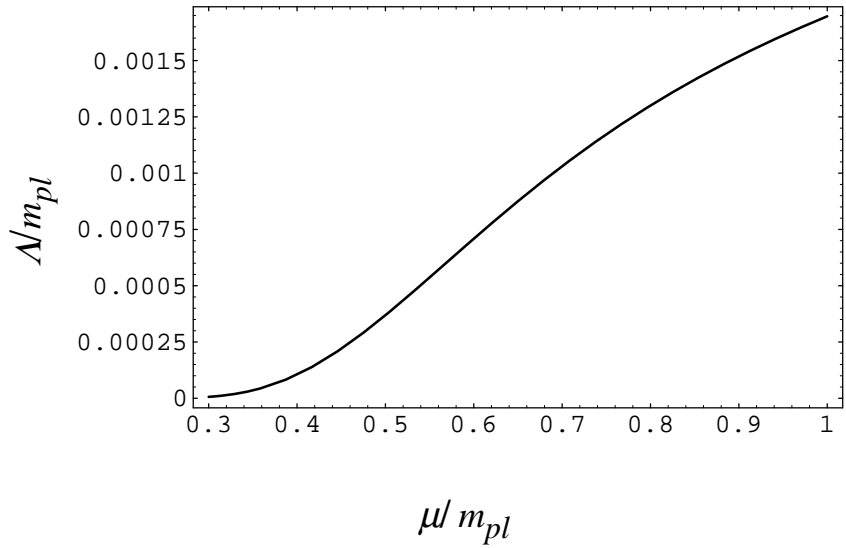

\caption{$\Lambda$ vs. symmetry breaking scale $\mu$ in natural inflation.}
\label{lambdavsmu}
\end{figure}

\begin{figure}
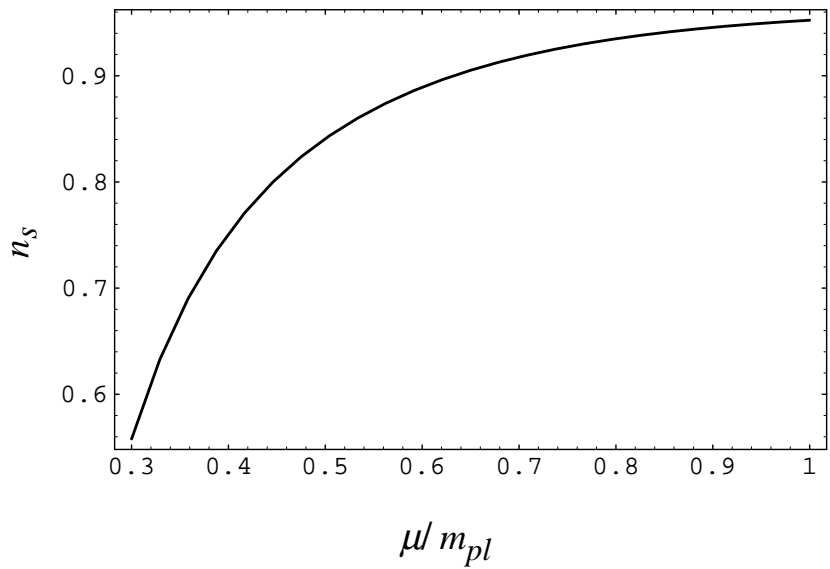

\caption{Spectral index $n_s$ vs. symmetry breaking scale $\mu$ in natural
inflation.}
\label{nsvsmu}
\end{figure}

We can constrain the scale $\Lambda$ as a function of $\mu$ by using the COBE
limit on the magnitude of density fluctuations (\ref{deltaCOBE}):
\begin{eqnarray}
\delta\left(\theta_i\right) =&& \left({2 \over \pi}\right)^{1/2}\left(\mu
\Lambda^2 \over m_{pl}^3\right) {\left[1 + \cos\left({\theta_i \over
\mu}\right)\right]^{3 \over 2} \over \sin\left({\theta_i \over \mu}\right)}
\left[1 - \epsilon\left(\theta_i\right) + \left(2 - \ln2 - \gamma\right)
\left(2 \epsilon\left(\theta_i\right) -
\eta\left(\theta_i\right)\right)\right]\cr \sim&&
10^{-5}\label{Lambdaconstraint}
\end{eqnarray}
where $\epsilon\left(\theta_i\right)$ and $\eta\left(\theta_i\right)$ are
independent of the scale $\Lambda$. As shown in Fig.\ \ref{lambdavsmu},
$\Lambda$ is highly sensitive to the original symmetry breaking scale: for
$\mu \sim m_{pl}$ we require $\Lambda \sim 10^{-3} m_{pl}$, and this drops
rapidly as $\mu$ decreases. The mass of the PNGB at the minimum of the
potential $\left(\theta / \mu\right) = \pi$ also decreases with $\mu$:
\begin{equation}M_{\theta}^2 \equiv V''\left[\left(\theta / \mu\right) =
\pi\right] = {\Lambda^4 \over \mu^2}\label{massoftheta}\end{equation}
We can constrain the value of the symmetry breaking scale $\mu$ using the
spectral index of density fluctuations (\ref{scalarindex}). Fig.\
\ref{nsvsmu} shows the spectral index $n_s$ as a function the of mass scale
$\mu$. The COBE result $n_s \geq 0.6$ then constrains the symmetry breaking
scale to $\mu \gtrsim 0.3\,m_{pl}$. We will take the upper limit on the mass
scale to be the Planck mass, $\mu = m_{pl}$, where the fluctuation power
spectrum is nearly scale-invariant, with $n_s = 0.95$.
A fluctuation spectrum consistent with cold dark matter (CDM) models requires
$n_s \sim 0.7$, and the corresponding symmetry breaking scale is $\mu =
0.36\,m_{pl}$. Natural inflation with $\mu \leq m_{pl}$ is not consistent
with a spectral index $n_s > 1$.

There is also the question of whether sufficient inflation will occur at all
with significant probability, especially for mass scales $\mu < m_{pl}$ where
the initial field values for which sufficient inflation occurs, $\theta \leq
\theta_i$, represent a very small portion of the available space of initial
conditions. As the universe cools to $T \sim \Lambda$, we expect the field
$\theta$ and its derivative $\dot\theta$ to take on different values in
different regions of the universe; here we will assume that the field is to a
good approximation uniform within any pre-inflation horizon volume. The
universe just prior to inflation then consists of a large number of causally
disconnected regions, each with independent initial conditions for $\theta$
and $\dot\theta$.
Each independent region will inflate a different amount, perhaps not at all,
depending on the conditions within that region. A successful model for
inflation is a model in which the {\it post}-inflation universe is strongly
dominated by regions in which $N\left(\theta\right) \geq 60$. It can be shown
that the initial value of $\dot\theta$ does not significantly affect the
number of e-folds of inflation\cite{knox93}, and hence we consider here only
the upper limit $\theta \leq \theta_i$ in (\ref{thetaupper}). Consider a
pre-inflation horizon volume $V_0$ and initial field value $\theta$: during
inflation, this region will expand to a volume $V = V_0 \exp\left[3
N\left(\theta\right)\right]$ The fraction of the volume of the post-inflation
universe for which $N\left(\theta\right) \geq 60$ is then \cite{adams93}
\begin{equation}F\left(N \geq 60\right) = 1 -
\left[\int_{\theta_i}^{\pi\mu}{\exp\left[3
N\left(\theta\right)\right]\,d\theta} \Bigg/
\int_{H/2\pi}^{\pi\mu}{\exp\left[3
N\left(\theta\right)\right]\,d\theta}\right]\end{equation}
Here a cutoff $H / 2 \pi$, the magnitude of quantum fluctuations on the scale
of a horizon size, has been introduced as the lower limit for the field
value. A successful inflationary theory then has the characteristic that
\begin{equation}\Pi\left(\theta_i\right) \equiv
\int_{\theta_i}^{\pi\mu}{\exp\left[3 N\left(\theta\right)\right]\,d\theta}
\Bigg/ \int_{H/2\pi}^{\pi\mu}{\exp\left[3
N\left(\theta\right)\right]\,d\theta} \ll 1\label{sufficient}\end{equation}
This condition is satisfied to a high degree for the range of symmetry
breaking scales allowed by observational constraints, with
$\Pi\left(\theta_i\right) < 10^{-62}$ for $\mu > 0.3\,m_{pl}$.

\section {Bounds on Parameters in Models with Potentials Generated by Fermion
Loop Corrections}
\label{models}

We now consider two specific models of natural inflation, both of which
generate the inflationary potential for the PNGB from fermion loop
corrections. Both models have three parameters: the mass scale $\mu$, the
bare fermion mass $m_0$ and a dimensionless Yukawa coupling $g$. As above, we
can constrain the scale $\mu$ in terms of the spectral index of the density
fluctuations created during inflation. For the purposes of this paper, we
will be taking $m_0 < \mu$. But now the COBE constraint on the amplitude of
density fluctuations (\ref{deltaCOBE}) results in a relation between the
Yukawa coupling and the fermion mass parameter $g = g(m_0)$. In the cases
considered here, the coupling $g$ {\it increases} with decreasing $m_0$, so
no fine-tuning of couplings is necessary.
In addition to the cosmological limits described above, we can also constrain
the models by requiring: (i) that the coupling constant remain perturbative,
$g < 1$, and (ii) that the PNGB have a nonzero mass at the minimum of the
potential, to avoid the presence of undesirable long-range forces in the
theory, and to make reheating possible. These constraints allow us to place
lower limits on $m_0$ in both cases.

\begin{figure}
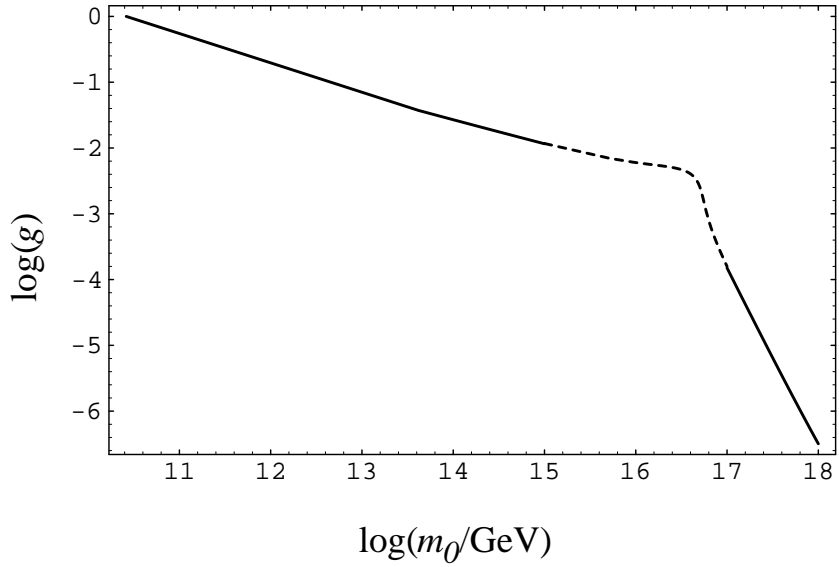

\caption{Coupling constant $g$ vs. $m_0$ with $\mu = m_{pl}$ (chiral model).
The solid lines are for the approximations $\left(m_0 / {g \mu}\right) \gg 1$
and $\left(m_0 / {g \mu}\right) \ll 1$. The dashed line is a numerical
solution for the exact potential (\protect\ref{Vchiral}) in the parameter
regime $\left(m_0 / {g \mu}\right) \sim O(1)$.}
\label{gvsmchiral}
\end{figure}

\begin{figure}
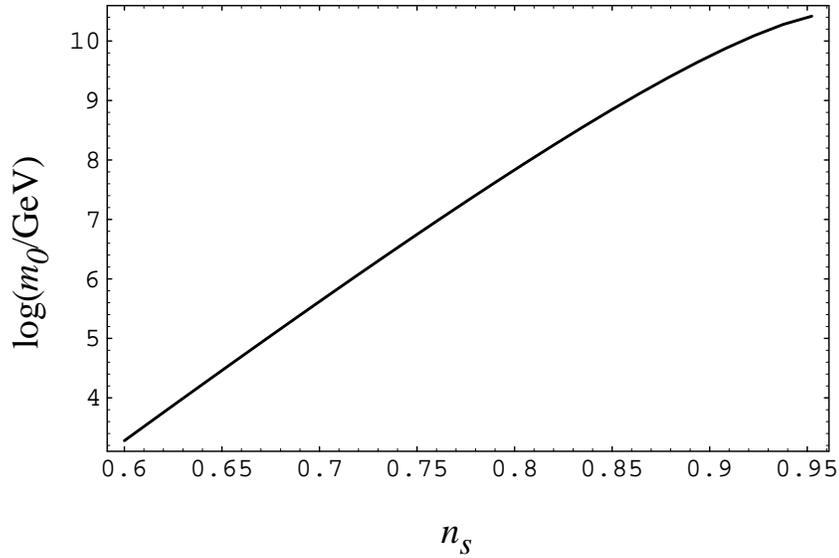

\caption{Lower limit on $m_0$  vs. spectral index $n_s$ (chiral model).}
\label{mvsnschiral}
\end{figure}

\begin{figure}
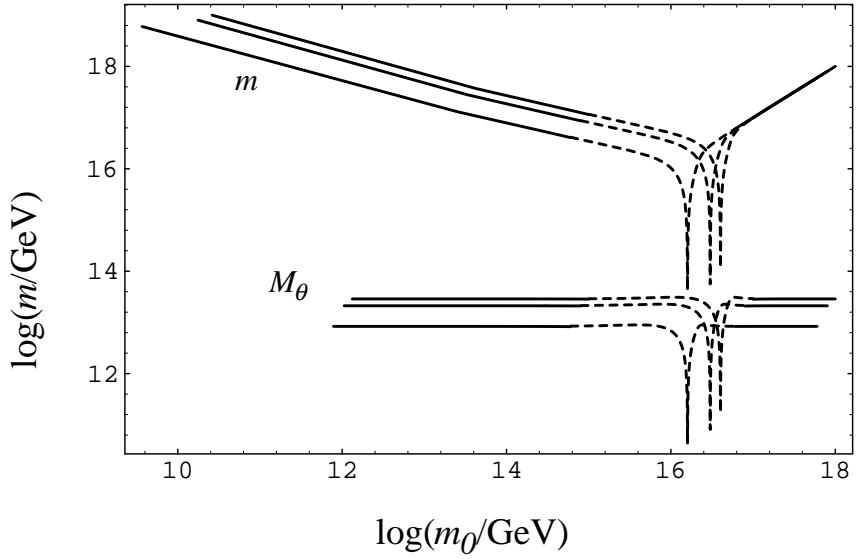

\caption{Fermion mass $m$ and PNGB mass $M_\theta$ vs. $m_0$ (chiral model),
with $\mu = 0.6 m_{pl}\mbox{, }0.8 m_{pl}\mbox{ and }m_{pl}$, where the
higher particle masses correspond to larger $\mu$. The solid lines are for
the limits $\left(m_0 / {g \mu}\right) \gg 1$ and $\left(m_0 / {g \mu}\right)
\ll 1$, and the dashed lines are for $\left(m_0 / {g \mu}\right) \sim O(1)$.}
\label{fermmassvsmchiral}
\end{figure}

{\it Chiral Fermions:}
Take a Lagrangian with a complex scalar field $\phi$ and fermions
{$\psi_{L,R} = {1\over2}(1 \pm \gamma^5) \psi$}, of the form\cite{adams93}:
\begin{eqnarray}
{\cal L} =&& {\cal L}_{kin} - {\bar\psi}_L m_0 \psi_R - {\bar\psi}_R m_0
\psi_L - g {\bar\psi}_L \phi \psi_R - g {\bar\psi}_R \phi^* \psi_L - \lambda
\left[ \phi^* \phi - \mu^2 \right]^2\cr =&& {\cal L}_{kin} - {\bar\psi}
m\left(\phi\right) \psi - \lambda \left[ \phi^* \phi - \mu^2 \right]^2\cr
{\cal L}_{kin} =&& \partial_\mu \phi^* \partial^\mu \phi + i \bar\psi
\gamma^\mu \partial_\mu \psi\label{chiralL}
\end{eqnarray}
The fields transform under a global $U(1)$ symmetry as
\begin{eqnarray}
\phi \rightarrow&& e^{i \alpha} \phi\cr \psi_L \rightarrow&& e^{i{\alpha
\over 2}} \psi_L \cr \psi_R \rightarrow&& e^{-i{\alpha \over 2}} \psi_R
\end{eqnarray}
so that the global $U(1)$ is exact in the limit $m_0 \rightarrow 0$. The
fermion mass term $- {\bar\psi}_L m_0 \psi_R - {\bar\psi}_R m_0 \psi_L$
explicitly breaks the $U(1)$ symmetry and results in the presence of a PNGB,
which will be the inflaton. Neglecting the one-loop shift in the vacuum
expectation value of the $\phi$ field, $<\phi> = \mu + \mbox{(one-loop
corrections)}$, we parameterize $\phi$ as in (\ref{adiabatic}), $\phi = \mu
\exp\left[{i \theta} / \mu\right]$.
The tree level fermion mass is
\begin{equation}m^2\left(\theta\right) \equiv m^\dagger\left(\theta\right)
m\left(\theta\right) = m_0^2 + g^2 \mu^2 + 2 g \mu m_0 \cos\left({\theta
\over \mu}\right)\label{ctreelevelmass}\end{equation}
Here we see that $m_0$ is the mass of the fermion at temperatures above the
scale of spontaneous symmetry breaking, $T > \mu$. In the spontaneously
broken phase, the physical fermion mass is given by (\ref{ctreelevelmass}),
and $m_0$ is treated as a parameter. The explicitly broken $U(1)$ symmetry is
reflected in the dependence of the fermion mass on the value of the PNGB
$\theta$. Quantum effects will generate a potential for $\theta$, which we
will use as an inflationary potential. The one-loop effective potential for
$\theta$ is \cite{weinberg73}
\begin{eqnarray}
V_1\left(\theta\right) = -&&{1 \over 64 \pi^2}
\left[m^2\left(\theta\right)\right]^2 \ln\left[ {m^2\left(\theta\right) \over
\mu^2} \right]\cr = -&&{1 \over 16 \pi^2} g^2 \mu^2 m_0^2 \left[ {1 \over 2}
\left({m_0 \over g \mu} + {g \mu \over m_0}\right) + \cos\left({\theta \over
\mu}\right) \right]^2 \times\cr &&\ln \left\lbrace 2 \left({g m_0 \over
\mu}\right) \left[ {1 \over 2} \left({m_0 \over g \mu} + {g \mu \over
m_0}\right) + \cos\left({\theta \over \mu}\right) \right] \right\rbrace
\label{Vchiral}
\end{eqnarray}
The minimum of $V_1\left(\theta\right)$ is at $\left({\theta / \mu}\right) =
\pi$, and the physical fermion mass at the minimum of the potential is
\begin{equation}
m\left[\left(\theta / \mu\right) = \pi\right] = \left\lbrace 2 g \mu m_0
\left[{1 \over 2} \left({m_0 \over g \mu} + {g \mu \over m_0}\right) -
1\right]\right\rbrace^{1 \over 2}
\label{cphysfermmass}
\end{equation}
The mass of the PNGB at the minimum of the potential is
\begin{eqnarray}
M^2_\theta \equiv&& V_1''\left[\left(\theta / \mu\right) = \pi\right]\cr =&&
- {1 \over 16 \pi^2} g^2 m_0^2 \left[ {1 \over 2} \left({m_0 \over g \mu} +
{g \mu \over m_0}\right) - 1\right] \left\lbrace 1 + 2 \ln\left[ 2 \left({g
m_0 \over \mu}\right) \left[ {1 \over 2} \left({m_0 \over g \mu} + {g \mu
\over m_0}\right) - 1 \right] \right] \right\rbrace
\label{cpseudomass}
\end{eqnarray}
Note that both the fermion mass (\ref{cphysfermmass}) and the PNGB mass
(\ref{cpseudomass}) vanish when $\left(m_0 / {g \mu}\right) = 1$, which we
exclude because a massless $\theta$ field will result in the presence of
undesirable long-range forces in the theory. In the limit of weak coupling,
$\left(m_0 / {g \mu}\right) \gg 1$, the effective potential becomes
\begin{equation}V_1\left(\theta\right) \simeq - {1 \over 32 \pi^2} g \mu
m_0^3 \left\lbrace 1 + 2 \ln\left[\left({m_0 \over
\mu}\right)^2\right]\right\rbrace \left[ 1 + \cos\left({\theta \over
\mu}\right)\right]\label{chiralweakcoup}\end{equation}
This is of the form (\ref{stdnat}), with
\begin{equation}\Lambda^4 = - {1 \over 32 \pi^2} g \mu m_0^3 \left\lbrace 1 +
2 \ln\left[\left({m_0 \over \mu}\right)^2\right]\right\rbrace\end{equation}
Note that this is a positive quantity for $\left({m_0 / \mu}\right) <
e^{-(1/4)}$. Remembering that $\Lambda$ is constrained in terms of the mass
scale $\mu$ by  (\ref{Lambdaconstraint}), this results in a relationship
between the Yukawa coupling $g$ and the fermion mass parameter $m_0$:
\begin{equation}g = - 32 \pi^2 \left({\Lambda \over \mu}\right)^4 \left({\mu
\over m_0}\right)^3 \left\lbrace 1 + 2 \ln\left[\left({m_0 \over
\mu}\right)^2\right]\right\rbrace^{-1}\end{equation}
The important feature to note is that the coupling constant $g$ increases
with decreasing $m_0$. In this limit, the physical fermion mass
(\ref{cphysfermmass}) is dominated by the parameter $m_0$:
\begin{equation}m\left[\left(\theta / \mu\right) = \pi\right] \simeq
m_0\end{equation}
In the limit of weak explicit symmetry breaking $\left(m_0 / {g \mu}\right)
\ll 1$, the effective potential (\ref{Vchiral}) becomes
\begin{equation}V_1\left(\theta\right) \simeq - {1 \over 32 \pi^2} g^3 \mu^3
m_0 \left[1 + 2 \ln\left(g^2\right)\right] \left[1 + \cos\left({\theta \over
\mu}\right)\right]\label{chiralstroncoup}\end{equation}
This is also of the form (\ref{stdnat}), with
\begin{equation}\Lambda^4 = - {1 \over 32 \pi^2} g^3 \mu^3 m_0 \left[1 + 2
\ln\left(g^2\right)\right]\label{chiralstrongcoupA}\end{equation}
The physical fermion mass (\ref{cphysfermmass}) is dominated by the symmetry
breaking scale $\mu$:
\begin{equation}m\left[\left(\theta / \mu\right) = \pi\right] \simeq g \mu
\end{equation}
In this case the coupling constant $g$ also increases with decreasing $m_0$,
so the necessity to fine-tune the coupling present in most inflationary
theories is entirely avoided. In fact, we have the opposite problem: for very
small $m_0$, the theory becomes strongly coupled and the perturbative
analysis above is inconsistent. Fig.\ \ref{gvsmchiral}
shows the coupling constant as a function of $m_0$ for the case $\mu =
m_{pl}$. We can place a constraint on $m_0$ by requiring that the Yukawa
coupling be perturbative, $g < 1$.  For $g > e^{- {1 / 4}}$, the right-hand
side of (\ref{chiralstrongcoupA}) becomes negative, which corresponds to
exchanging the location of the maxima and minima of the potential
(\ref{Vchiral}). Thus, for $g < 1$, we have
\begin{equation}\Lambda^4 = \left|{1 \over 32 \pi^2} g^3 \mu^3 m_0 \left[1 +
2 \ln\left(g^2\right)\right]\right|\label{chiralstroncoupB}\end{equation}
Setting $g = 1$ in (\ref{chiralstroncoupB}), the lower limit on $m_0$ is then
\begin{equation}m_0 \geq 32 \pi^2 \mu \left({\Lambda \over \mu}\right)^4
\label{chiralmasslimit}\end{equation}
This is a limit based on theoretical consistency and not on phenomenology.
However, for $m_0$ below this limit, the entire procedure of calculating the
inflationary potential (\ref{Vchiral}) in perturbation theory is seen to be
invalid. For a scale-invariant power spectrum, $n_s \sim 1$, we require a
symmetry breaking scale near the Planck mass, $\mu \sim m_{pl}$. In this
case, the lower limit on $m_0$ is $m_0 \gtrsim 10^{10} \mbox{GeV}$. For a CDM
power spectrum, $n_s = 0.7$, the lower limit drops to $m_0 \gtrsim\,10^5
\mbox{GeV}$. For the lowest value of $n_s = 0.6$ allowed by COBE, $\mu =
0.32\,m_{pl}$, the lower limit is $m_0 \gtrsim 10^3 \mbox{GeV}$.
However, for very small values of $m_0$, the hierarchy of scales $m_0 \ll
\mu$ will not in general be maintained when the parameters are renormalized,
and one must invoke an additional symmetry of the Lagrangian to preserve the
difference in scales. Ref. \cite{adams93} contains a description of one
possible mechanism. Fig.\ \ref{mvsnschiral} shows the lower limit on $m_0$ as
a function of the spectral index.
We can also calculate the fermion mass $m$ (\ref{cphysfermmass}) and the PNGB
mass (\ref{cpseudomass}) as a function of $m_0$ (Fig.\
\ref{fermmassvsmchiral}). Typical fermion masses are in the range $10^{16}
\mbox{-} 10^{19} \mbox{GeV}$, with PNGB masses in the range $10^{13} \mbox{-}
10^{14} \mbox{GeV}$. For all values of the parameters, the fermion mass is
greater than the mass of the PNGB, so that the decay $\theta \rightarrow
\bar\psi\ \psi$ is forbidden, and reheating must take place in some other,
unspecified, sector of the theory.

We note that, although we obtain a potential of the form (\ref{stdnat}) in
both limits $\left(m_0 / {g \mu}\right) \gg 1$ and $\left(m_0 / {g
\mu}\right) \ll 1$, the values obtained for the parameter $\Lambda$
\begin{eqnarray}
\Lambda^4&& \sim g \mu m_0^3\ \ \mbox{for}\ \ \left({m_0 / g \mu}\right) \gg
1\cr
\Lambda^4&& \sim g^3 \mu^3 m_0\ \ \mbox{for}\ \ \left({m_0 / g \mu}\right)
\ll 1
\end{eqnarray}
differ from the result of $\Lambda^4 \sim \left(g \mu m_0\right)^2$ in Ref.
\cite{adams93}.

\begin{figure}
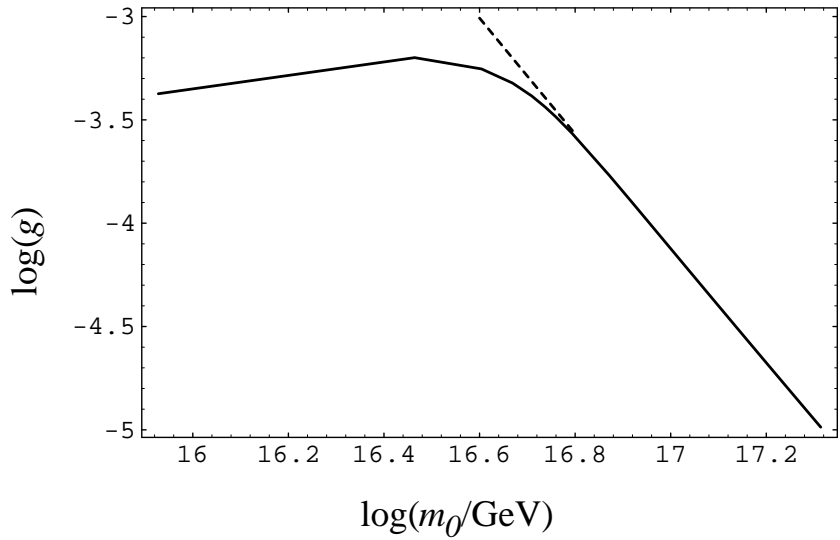

\caption{Coupling constant $g$ vs. $m_0$, with $\mu = m_{pl}$ (non-chiral
model). The solid line is the result for the exact potential
(\protect\ref{Vnonchiral}), and the dashed line is the result for the limit
$\left({m_0 / {2 g \mu}}\right) \gg 1$.}
\label{gvsmnonchiral}
\end{figure}

\begin{figure}
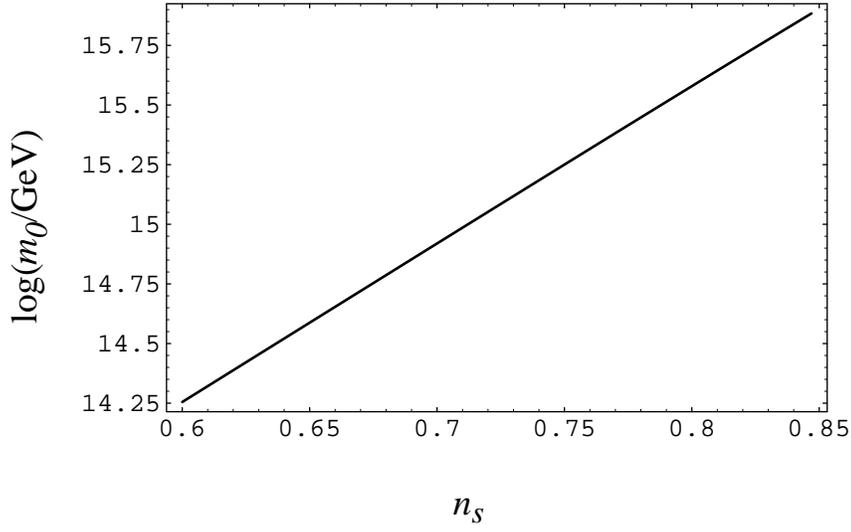

\caption{Lower limit on $m_0$ vs. spectral index $n_s$ with $\left({m_0 / {2
g \mu}}\right) = 1$ (non-chiral model).}
\label{mvsnsnonchiral}
\end{figure}

{\it Non-Chiral Fermions:}
We now construct a model in which the global $U(1)$ is preserved by the
fermion mass term, but broken by the Yukawa coupling:
\begin{equation} {\cal L} = {\cal L}_{kin} - {\bar\psi} m_0 \psi - g
{\bar\psi} \phi \psi - g {\bar\psi} \phi^* \psi - \lambda \left[ \phi^* \phi
- \mu^2 \right]^2 \label{nonchiralL}\end{equation}
The fields transform under a global $U(1)$ symmetry as $ \phi \rightarrow
e^{i \alpha} \phi$, $\psi \rightarrow e^{i \alpha} \psi$, so that the $U(1)$
symmetry is exact when $g \rightarrow 0$. The Yukawa couplings explicitly
break the $U(1)$ symmetry and result in the presence of a PNGB, which will be
the inflaton. Parameterizing $\phi$ as in (\ref{adiabatic}), $\phi = \mu
\exp\left[{i \theta} / \mu\right]$, the tree level fermion mass is
\begin{equation}m^2\left(\theta\right) = \left[m_0 + 2 g \mu
\cos\left({\theta \over \mu}\right)\right]^2\label{ncphysfermmass}
\end{equation}
The one-loop effective potential for the PNGB is
\begin{eqnarray}
V_1\left(\theta\right) =&& -{1 \over 64 \pi^2}
\left[m^2\left(\theta\right)\right]^2 \ln\left[ {m^2\left(\theta\right) \over
\mu^2} \right]\cr
=&& - {1 \over 4 \pi^2} g^4 \mu^4 \left[ \left({m_0 \over 2 g \mu}\right) +
\cos\left({\theta \over \mu}\right) \right]^4 \ln\left\lbrace 4 g^2 \left[
\left({m_0 \over 2 g \mu}\right) + \cos\left({\theta \over \mu}\right)
\right]^2 \right\rbrace \label{Vnonchiral}
\end{eqnarray}
When $\left({m_0 / 2 g \mu}\right) \leq 1$, the effective potential has two
equivalent minima
\begin{equation}\left({\theta_{min} \over \mu}\right) = \cos^{-1}
\left[-\left({m_0 \over 2 g \mu}\right)\right]\end{equation}
However, the PNGB is {\it massless} in the true vacuum, as
$V''\left(\theta_{min}\right) = 0$. We must therefore exclude this parameter
range, since there will be long-range forces associated with the field
$\theta$.
In the parameter range $\left({m_0 / 2 g \mu}\right) > 1$, the potential
$V_1\left(\theta\right)$ has a single minimum at $\left({\theta / \mu}\right)
= \pi$. The mass of $\theta$ at the minimum of the potential is:
\begin{equation}
M_{\theta}^2 \equiv V''\left(\theta = \pi \mu\right) = - {1 \over 2 \pi^2}
g^4 \mu^2 \left[\left({m_0 \over 2 g \mu}\right) - 1\right]^3 \left\lbrace 1
+ 2 \ln\left(4 g^2 \left[\left({m_0 \over 2 g \mu}\right) -
1\right]^2\right)\right\rbrace
\label{noncpseudomass}
\end{equation}
The physical fermion mass is
\begin{equation}
m\left[\left(\theta / \mu\right) = \pi\right] = 2 g \mu \left[\left({m_0
\over 2 g \mu}\right) - 1\right]
\label{noncphysfermmass}
\end{equation}
We can simplify the form of the potential in the limit where $M_{\theta}$ is
large:
\begin{equation} \left({m_0 \over 2 g \mu}\right) \gg 1
\label{weakbreak}\end{equation}
Introducing a cosmological constant so that $V_1\left(\theta = \pi \mu\right)
= 0$, the potential becomes
\begin{equation} V_1\left(\theta\right) = {1 \over 16 \pi^2} g \mu m_0^3
\left[ 4 \ln\left({\mu \over m_0}\right) - 1 \right] \left[ 1 +
\cos\left({\theta \over \mu}\right) \right]\end{equation}
This is of the form (\ref{stdnat}), with
\begin{equation} \Lambda^4 = {1 \over 16 \pi^2} g \mu m_0^3 \left[ 4
\ln\left({\mu \over m_0}\right) - 1 \right] \end{equation}
In this limit, the physical fermion mass is dominated by the parameter $m_0$:
\begin{equation}m\left[\left(\theta / \mu\right) = \pi\right] \simeq
m_0\end{equation}
We again have a relationship between the Yukawa coupling $g$ and $m_0$:
\begin{equation} g = 16 \pi^2 \left({\Lambda \over \mu}\right)^4 \left({\mu
\over m_0}\right)^3 \left[ 4 \ln\left({\mu \over m_0}\right) - 1 \right]^{-1}
\end{equation}
In this limit, the mass of $\theta$ is, as usual, $M_{\theta}^2 = (\Lambda^4
/ \mu^2)$. Again, successful inflation requires the coupling to become
stronger with decreasing $m_0$.  Note, however, that the approximation
(\ref{weakbreak}) breaks down for $g \sim 1$, $m_0 < \mu$, and we cannot use
the requirement of small coupling to limit $m_0$. In addition, successful
inflation occurs in a significantly large parameter space which does not
correspond to the limiting case (\ref{stdnat}).
We must consider the case $\left({m_0 / 2 g \mu}\right) \sim O(1)$
separately, since it requires solution of the inflationary constraints for
the exact potential (\ref{Vnonchiral}). Numerical solution of the
inflationary constraints indicates that the theory never becomes strongly
coupled, and in fact $g$ ``turns over'' and begins to decrease for small
$m_0$ (Fig.\ \ref{gvsmnonchiral}). However, the mass of $\theta$ drops from
its value in the limit (\ref{weakbreak}), $M_{\theta}^2 = (\Lambda^4 /
\mu^2)$, to zero as $\left({m_0 / 2 g \mu}\right)$ approaches $1$. The
requirement that $M_{\theta}$ be greater than zero allows us to place a lower
limit on $m_0$.
Taking $\left({m_0 / 2 g \mu}\right) = 1$, the potential is
\begin{equation}V_1\left(\theta\right) = - {1 \over 4 \pi^2} g^4 \mu^4 \left[
1 + \cos\left({\theta \over \mu}\right) \right]^4 \ln\left\lbrace 4 g^2
\left[ 1 + \cos\left({\theta \over \mu}\right) \right]^2
\right\rbrace\end{equation}
The field value at the end of inflation $\theta_f$ is defined in terms of the
first-order slow-roll parameter $\epsilon\left(\theta\right)$:
\begin{equation}\epsilon\left(\theta_f\right) \equiv {m_{pl}^2 \over 16 \pi}
\left[{V_1'\left(\theta_f\right) \over V_1\left(\theta_f\right)}\right]^2 =
1\end{equation}
where
\begin{equation}{m_{pl} V_1'\left(\theta\right) \over V_1\left(\theta\right)}
= - 2 \left({m_{pl} \over \mu}\right) \left[ {\sin\left({\theta \over
\mu}\right) \over 1 + \cos\left({\theta \over \mu}\right)} \right] \left[ {1
+ 2 \ln\left\lbrace 4 g^2 \left[ 1 + \cos\left({\theta \over \mu}\right)
\right]^2 \right\rbrace \over \ln\left\lbrace 4 g^2 \left[ 1 +
\cos\left({\theta \over \mu}\right) \right]^2 \right\rbrace} \right]
\end{equation}
For a small coupling $g$, $\epsilon\left(\theta\right)$ is approximately
\begin{equation}\epsilon\left(\theta\right) \simeq {1 \over \pi}
\left({m_{pl} \over \mu}\right)^2 {\sin^2\left({\theta \over \mu}\right)
\over  \left[ 1 + \cos\left({\theta \over \mu}\right)
\right]^2}\end{equation}
Then for $\epsilon\left(\theta_f\right) = 1$
\begin{equation}\cos\left({\theta_f \over \mu}\right) = {1 - \pi \left(\mu /
m_{pl}\right)^2 \over 1 + \pi \left(\mu / m_{pl}\right)^2}\end{equation}
The upper limit on the initial field value $\theta \leq \theta_i$ is defined
such that the number of e-folds of inflation $N\left(\theta_i\right) = 60$,
where
\begin{equation}N\left(\theta\right) \equiv {8 \pi \over m_{pl}^2}
\int_{\theta_f}^{\theta_i}{{V_1\left(\theta\right) \over
V_1'\left(\theta\right)}\,d\theta} = 2 \pi \left({\mu \over m_{pl}}\right)^2
\ln \left[ {1 - \cos\left({\theta_f \over \mu}\right) \over 1 -
\cos\left({\theta_i \over \mu}\right) }\right]\end{equation}
Taking $\theta_i$ such that $N\left(\theta_i\right) = 60$,
\begin{equation}\cos\left({\theta_i \over \mu}\right) = 1 -  \left[{2 \pi
\left(\mu / m_{pl}\right)^2 \over 1 + \pi \left(\mu / m_{pl}\right)^2}\right]
\exp\left[-{30 \over \pi} \left({m_{pl} \over
\mu}\right)^2\right]\end{equation}
The coupling constant $g = (m_0 / 2 \mu)$ can then be determined as a
function of the symmetry breaking scale $\mu$ by numerical solution of the
COBE constraint (\ref{deltaCOBE}). For $\mu \sim m_{pl}$, the spectral index
is $n_s \simeq 0.85$ and the lower limit on $m_0$ is $m_0 \gtrsim 10^{16}
\mbox{GeV}$. We get $m_0 \gtrsim 10^{15} \mbox{GeV}$ for the CDM power
spectrum, $n_s = 0.7$. For the lowest value of $n_s = 0.6$ allowed by COBE,
$m_0 \gtrsim 10^{14} \mbox{GeV}$. Fig.\ \ref{mvsnsnonchiral} shows the lower
limit on $m_0$ as a function of the spectral index $n_s$.

To obtain a spectral index $n_s > 0.85$, we must take $\left(m_0 / {2 g
\mu}\right) > 1$. The nearly scale-invariant limit, $n_s = 0.95$, occurs for
$m_0 \gtrsim 10^{17} \mbox{GeV}$.

\begin{figure}
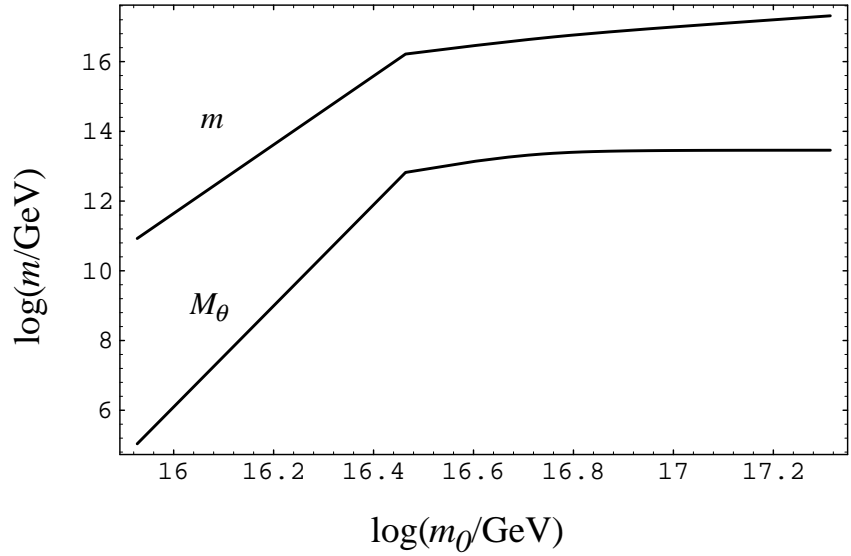

\caption{Fermion mass $m$ and PNGB mass $M_\theta$ vs. $m_0$, for the case
$\mu = m_{pl}$ (non-chiral model).}
\label{fermmassvsmnonchiral}
\end{figure}

Fig. \ref{fermmassvsmnonchiral} shows the fermion mass $m$
(\ref{noncphysfermmass}) and the PNGB mass $M_\theta$ (\ref{noncpseudomass})
as functions of the parameter $m_0$. Here, as in the chiral model, the PNGB
is lighter than the fermions for all values of the parameters.

\section {Conclusions}
\label{conclusions}

We have considered two models for inflation in which a PNGB acquires a
potential as a result of fermion loop corrections in a Lagrangian with an
explicitly broken global $U(1)$ symmetry. Although the $U(1)$ symmetry
considered in these models is very simple, the mechanisms of the explicit
symmetry breaking are quite general. In both models, successful inflation
occurs without fine-tuning of parameters, and the potential (\ref{stdnat})
arises as a limiting case of the exact one-loop effective potential for the
PNGB. In the case of the chiral model, our result
$\Lambda^4 \sim g \mu m_0^3\ \mbox{for}\ \left({m_0 / g \mu}\right) \gg 1\
\mbox{and}\
\Lambda^4 \sim g^3 \mu^3 m_0\ \mbox{for}\ \left({m_0 / g \mu}\right) \ll 1$
differs from the result of $\Lambda^4 \sim \left(g \mu m_0\right)^2$ in Ref.
\cite{adams93}. In the case of the non-chiral model, successful inflation
occurs in parameter regimes for which the limiting case (\ref{stdnat}) is not
applicable.

It is also natural to ask similar questions about models in which the
explicit symmetry breaking takes place in the gauge sector rather than the
fermi sector. This is the subject of continuing work.

This work is supported in part by U.S. Department of Energy Grant No.
DEFG-ER91-406672.


\end{document}